\documentstyle[aps,prl,preprint,epsf]{revtex}
 \def\half{\mbox{$1\over 2$}}

 \def\beq{\begin{equation}}
 \def\eeq{\end{equation}}
 \def\beqa{\begin{eqnarray}}
 \def\eeqa{\end{eqnarray}}
 
 \def\LP{\left(}
 \def\RP{\vphantom{\half} \right)}

 \gdef\aver#1{\left\langle #1 \right\rangle}
 \gdef\s#1{\! #1 \!}
 \gdef\l#1{\> #1 \>}
 \gdef\Eq#1{Eq.~(\ref{#1})}
 
 \gdef\vec#1{{\bf #1}}

\begin{document}
\draft

\preprint{NSF-ITP-96-64}

\title{
Random Scattering Matrices \\
and the Circuit Theory of Andreev Conductances }

\author{Nathan Argaman}
\address{ Institute for Theoretical Physics,
University of California, Santa Barbara, CA 93106, USA }

\date{August '96}

\maketitle

\begin{abstract}

The conductance of a normal-metal mesoscopic system in proximity
to superconducting electrode(s) is calculated.
The normal-metal part may have a general geometry, and is described as
a ``circuit'' with ``leads'' and ``junctions''.   The junctions are
each ascribed a scattering matrix which is averaged over the circular
orthogonal ensemble, using recently-developed techniques.
The results for the electrical conductance reproduce and extend 
Nazarov's circuit theory, thus bridging between the scattering and the 
bulk approaches.  The method is also applied to the heat conductance.

\end{abstract}

\pacs{PACS numbers: 74.50.+r, 74.80.Fp, 73.23.Ps}

Mesoscopic normal-metal--superconductor ({N-S}) proximity
effects are often described theoretically using the Usadel equations
\cite{Usadel} --- the partial differential equations of inhomogeneous
(dirty) superconductivity.  However, the ohmic resistance of the
normal-metal part of the structure \cite{review} can also be found
from the ``circuit theory of Andreev conductance'' --- a set of
algebraic relations recently derived from the Usadel equations by Yuli
Nazarov \cite{Nazarov} (the Usadel equations themselves are derived
from the Eilenberger equations of clean superconductors, which
ultimately derive from the Keldysh technique).  Here we pursue an
alternative approach, which associates a random  scattering matrix
with each of the junctions in the ``circuit''.  These matrices are to
be averaged over one of the circular ensembles of random matrix
theory, employing diagrammatic techniques which were developed
specifically for such purposes \cite{AZ,BB}.

Electron and hole excitations can enter the normal-metal part 
of such {N-S} structures (the ``circuit'') from one of the normal
electrodes ({N}), but at low excitation energies they can not
enter into the superconducting electrodes ({S}) because of the
gap.  Instead, Andreev reflections may occur --- an electron
may evolve into a Cooper pair in {S}, leaving behind a hole in 
the Fermi sea of the normal-metal part \cite{Andreev}, with a
wavevector related to that of the impinging electron by time-reversal.  
We will assume for simplicity, as did Nazarov \cite{Nazarov}, that the
applied voltages and the ambient temperatures are very small, so that
the energy dependence of the scattering properties of the structure
can be neglected, and the quantum mechanical propagation is fully
coherent.  We also assume the absence of a magnetic field (except for
Aharonov-Bohm phases), the absence of electron-electron interactions
in the normal part of the structure, and unbroken time-reversal
symmetry and spin degeneracy.

Unaveraged quantum mechanical transport in a wide variety of
structures may be described \cite{Been-old} using two unitary
matrices: $G_0$ represents the ``non-local'' transport in the ``leads'',
and $S$ represents random scattering within each of the ``junctions'',
as well as Andreev scattering at the external superconducting electrodes.  
We denote the number of internal junctions in the system by $K$, and
the number of junctions and external electrodes by $M$ (we have at
least one {N} and one {S} electrode, so $M \geq K+2$; see Fig.~1a).
The number of transverse modes in the lead connecting junctions $k$
and $m$ is denoted by $W_{k,m}$, so that $N_k=\sum_m W_{k,m}$ is the
size of the scattering matrix of the $k$th junction.  The $N_k$s and
the (nonzero) $W_{k,m}$s are treated as large parmeters, and the
conductances are evaluated to leading order --- weak localization and
conductance fluctuation corrections will not be considered here.  The
matrices $G_0$ and $S$ are of size $2N_{\rm tot} \times 2N_{\rm tot}$,
where $N_{\rm tot} = \sum_{m=1}^M N_m$ is the total number of modes,
and the factor of 2 represents the electron-hole subspace.  $G_0$ is
non-diagonal in the $M \times M$ block structure, but diagonal in the
electron-hole space.  $S$ is block-diagonal, with $K$
diagonal blocks consisting of a random scattering matrix for
the electron-electron sub-block, and the complex conjugate of that
matrix for the hole-hole sub-block.  The remaining diagonal blocks of
$S$ either vanish, for an external normal electrode, or are off-diagonal
in the electron-hole space: the electron-hole Andreev-scattering 
sub-block for a superconducting electrode $m$ is equal to 
$i \exp(-i \chi_m)$ (times a unit $N_m \times N_m$ matrix), and the
hole-electron sub-block is $i \exp(i \chi_m)$.  Here the $\chi_m$s are
the given, time-independent phases of the superconducting order
parameter in $m$.  The extra factors of $i$ lead
to {\it destructive} interference after two consecutive Andreev
scatterings, and are responsible for, e.g., proximity-induced gap-like
features in the normal part of the structure.

We now average the $K$ random-scattering diagonal blocks of $S$ over
the appropriate matrix ensemble --- the circular orthogonal ensemble
(COE), justifying our expressions both physically and technically (see
Ref.~\cite{AZ} for details of the latter).  Diagrams for the
conductance may be drawn as in Fig.~1b, with matrix multiplication
implied along the bottom and the top of the diagram.  The matrices
along the top line are hermitean conjugated, so that the diagrams
represent probabilities, or amplitudes multiplied by complex
conjugates of amplitudes.  Averaging is evaluated to leading order by 
connecting the random elements of $S$, represented as double lines, in
all possible ways in the plane of the diagram without crossing lines.
These connections or couplings (Fig.~1c) join together $n$ occurrences
of the $k$th random scattering matrix and $n$ occurrences of its hermitean
(or complex) conjugate, in alternating order; the corresponding weights 
are $(-1)^{n-1} c_n / N_k^{2n-1}$ with $c_n = (2n)!/2(2n-1)(n!)^2$ 
(these are deduced in Refs.~\cite{AZ} and \cite{BB} from unitarity).
Each $2n$-fold coupling divides the plane of the diagram into $2n$
regions, with matrix multiplication performed independently around the
periphery of each region.  The blocks of $S$ which are not
averaged over --- those representing Andreev scattering at the
superconducting electrodes --- are drawn as black semicircles.

The averaged propagator $G = \aver{ G_0 + G_0 S G_0 + \dots }$ is
written as $G = G_0/(I-\Sigma G_0)$, where the ``one particle
irreducible part'', or ``self-energy'' $\Sigma$ represents the
``average'' scattering by the junctions (Fig.~1d).  The structure
of $\Sigma$ is particularly simple, because (a) both normal scattering
and off-diagonal (in mode space) electron-hole scattering average to zero 
due to the random phases involved, and (b) the diagonal electron-hole
scattering elements of $\Sigma$ must be equal to each other within
each of the $K$ random-scattering blocks, due to ergodicity in each
junction.  We denote the electron-hole averaged Andreev scattering
amplitude by $if_k$, where $f_k$ is a complex number in the unit circle.  
The hole-electron amplitude is then $if_k^*$.  Direct Andreev scattering
(black semicircles) is also conveniently included in $\Sigma$: for the
superconducting electrodes $f_m=\exp(-i\chi_m)$, and for the normal ones 
$f_m=0$.

The behavior of the amplitudes $f_k$ is similar to that of voltages
--- if a certain junction is connected by a lead with a high
conductance to an electrode with a given value of $f_m$, then the
value of $f_k$ in that junction will tend towards $f_m$.  These
amplitudes are physically observable in the following sense: if we
were to inject electrons directly into the $k$th junction from an
additional normal electrode which is relatively weakly coupled to that
junction --- a ``noninvasive voltage probe'' --- we would find a
strong beam of holes being retro-reflected into that voltage probe,
with an intensity of $|f_k|^2$ times that of the electron beam.  The
fact that these amplitudes are of order 1, rather than of order
$1/N_k$, has been referred to in the literature as the 
``giant (Andreev) backscattering peak'' \cite{Been_GBP}.  It can be 
understood on the basis of the analogy between Andreev reflections and
phase-conjugating mirrors in optics.

For the specific case of ideal leads, $G = G_0 + G_0 \Sigma G_0 + \dots$ 
is a simple sum over multiple Andreev reflections.  For example, 
$G_{kh,j;ke,j}^m = if_m+if_mif_k^*if_m+\dots = if_m/(1+f_mf_k^*)$, 
where $j$ is the index of one of the $W_{k,m}$ modes in the lead
connecting $k$ and $m$ (the superscript $m$ has been added to
emphasize this; $kh$ and $ke$ denote the $k$th diagonal electron-hole 
sub-block).  On the other hand, the diagrams for $\Sigma$ express
$f_k$ as a sum of powers of $\alpha_k$ and $\alpha_k^*$, where 
$i \alpha_k = (1/N_k) \sum_{j=1}^{N_k} G_{kh,j;ke,j}$ is a
trace of a sub-block of $G$.  When the explicit values of the $c_n$s 
are used and the relationship between $\alpha_k$ and $f_k$ is
inverted, it reads simply
\beq \label{alpha}
 { f_k \over 1+|f_k|^2 }  \l=  \alpha_k  \l=  
{1 \over N_k} \sum_m W_{k,m} \, {f_m \over 1 + f_m f_k^*}  \; .
\eeq 
This is consistent with the physical requirement that $f_k=f_m$ if 
all the $f_m$s in the sum are equal to each other.  \Eq{alpha} may be
rewritten, using $\sum_m W_{k,m} = N_k$, as an expression of
``spectral current conservation'':
\beq \label{Jdef}
\sum_m J_{k,m}  =  0  \quad ; \quad 
J_{k,m}  = 
2 \sum_{j=1}^{W_{k,m}} \LP (1 \s+ |f_k|^2) G_{kh,j;ke,j}^m - f_k \RP 
=  W_{k,m} \, 2 {f_m - f_k \over 1 + f_m f_k^*}  \; ,
\eeq
where $J_{k,m}$ denotes the dimensionless (complex) ``spectral
current'' in the $(k,m)$ lead. 

The ``two-particle irreducible vertex'', $\Gamma$ is represented
diagrammatically in Fig.~1e.  It is also block-diagonal, with all of
the elements of the $k$th block given by  
${1+|f_k|^2 \over N_k (1 - |f_k|^2)}
{\quad 1 \quad -|f_k|^2 \choose -|f_k|^2 \quad 1 \quad}$ (each of the
four sub-blocks are full matrices with identical elements).
This expression is the only one
consistent with unitarity or current conservation \cite{rem}.  
The two-particle propagator, or ``diffuson'' is  given by 
$D = \Gamma + \Gamma t \Gamma + \dots$, where the elements of $t$ 
are equal to the absolute squares of the corresponding elements of $G$.
Rather than evaluating the matrix $D$ directly, as would be necessary
in order to express the conductances in the usual scattering approach
\cite{Been-old,cmtrx}, we define occupation probabilities $v$ for the
electron and hole modes leaving any of the junctions or electrodes.
For example,
$v_{me(h),j}(\epsilon)= 1/\LP 1+\exp[(\epsilon\pm eV_m)/k_BT_m] \RP$
for an external normal electrode $m$ with an applied 
voltage $V_m$ and a temperature $T_m$
(here $\epsilon$ is the excitation energy relative to the Fermi
surface of the superconducting electrodes).  
The values of $v$ for modes leaving internal junctions of the circuit 
are given by $v = \Gamma t v$, where the probabilities $v$ on the
right hand side represent the neighbouring junctions or electrodes,
and the factors of $t$ and $\Gamma$ represent transport into and
inside the junction, respectively.  It is convenient to rewrite this as
\beq \label{vGtv}
D^{-1} v = ( \Gamma^{-1} - t) v = 0 \; ,
\eeq
with the equality required only for the $K$ internal junctions.

As we assume that no inelastic processes occur in the structure, 
the electric current $I$ and the quasiparticle current $Q$ are
conserved differentially at each junction $k$:
\beq \label{curc}
\sum_m I_{k,m}(\epsilon) = 0  \quad ; \quad 
\sum_m Q_{k,m}(\epsilon) = 0  \; .
\eeq
The total electric current, 
$I_{k,m}= \int_0^\infty d\epsilon \> I_{k,m}(\epsilon)$, 
and heat current,
$Q_{k,m}= \int_0^\infty \epsilon \> d\epsilon \> Q_{k,m}(\epsilon)$, 
are consequently conserved as well.  For ideal leads, \Eq{vGtv}
implies that \cite{rem}
\beqa \label{Idef}
I_{k,m} & = & {2 e^2 \over h}
W_{k,m} {(1+|f_m|^2) (1+|f_k|^2) \over |1+f_m f_k^*|^2} (V_k - V_m) 
\; ;  \\ \label{Qdef}
Q_{k,m} & = &  {2 \over h}
W_{k,m} {(1-|f_m|^2) (1-|f_k|^2) \over |1+f_m f_k^*|^2} (U_k - U_m)
\; ,
\eeqa
where the voltages $V_m$ and the excitation-energy densities $U_m$ 
are defined by \cite{imbalance}
\beq \label{Vdef}
V_m  \l=  {1 \over e} \int_0^\infty d\epsilon \> 
{v_{hm}-v_{em} \over 1 + |f_m|^2}  \quad ; \quad
U_m  \l=  \int_0^\infty d\epsilon \; \epsilon \> 
{v_{hm}+v_{em} \over 1 - |f_m|^2}  \; .
\eeq
These definitions are reasonable because $|f_m|=0$ at the {N} 
electrodes.  At the {S} electrodes, both the voltages $V_m$ and 
the heat currents $Q_{k,m}$ vanish.  
The net heat current into an external normal electrode $m$ is
$Q_{k,m} - I_{k,m} V_m$ and not just $Q_{k,m}$, because of the shift 
in the Fermi level. 
This completes the derivation for ideal leads ---  
the conservation of the currents of Eqs.~(\ref{Jdef}), (\ref{Idef})
and (\ref{Qdef}) at each junction determines the Andreev amplitudes
$f_k$, the voltages $V_k$ and the energy densities $U_k$
self-consistently.  Although seemingly local, this scheme takes into
account the long-range proximity effects.  Aharonov-Bohm effects can
also be included simply by gauging out the vector potential $\vec A$
separately along each lead, i.e.\ by multiplying each occurence of
$f_m$ in the equations by 
$\exp(i2e\int_m^k \vec A \s\cdot d \vec x / \hbar)$.

Electronic transport in a non-ideal $(k,m)$ lead may be described by
an arbitrary (but non-random) scattering matrix.  Using the polar
decomposition of this matrix, and the symmetry of the COE, the
transmission and reflection eigenvectors can be absorbed into the
scattering of the $k$ and $m$ junctions (assuming that time-reversal
symmetry is preserved).  This
means that without loss of generality we may associate a transmission
probability $T_j$ with each of the $W_{k,m}$ modes in the lead.  The
corresponding transmission and reflection amplitudes in $G_0$ are
equal to $i\sqrt{T_j}$ and $\sqrt{1-T_j}$ respectively.  Expressing
$G= G_0+ G_0 \Sigma G_0 + \dots$ in terms of the $T_j$s, $f_k$ and
$f_m$ involves only inverting a $j$-dependent $2 \times 2$ matrix [the
electron-hole structure is simplified by separating the even (Andreev
reflecting) terms in the sum from the odd (non-Andreev) terms].  With
the notation $\beta_{k,m} = (f_m \s- f_k)/(1 \s+ f_m f_k^*)$, 
the spectral current becomes
\beq \label{Jgen}
J_{k,m}  \l=  2 \sum_{j=1}^{W_{k,m}}  
T_j \beta_{k,m} \LP 1+ (1-T_j) |\beta_{k,m}|^2 \RP^{-1} \; .
\eeq
Expressions for charge and heat transport are similarly obtained, 
by replacing the factor $W_{k,m}/|1+f_m f_k^*|^2$ in 
Eqs.~(\ref{Idef}) and (\ref{Qdef}) 
with $\sum_{j=1}^{W_{k,m}} ( t_{ke,j;me,j} \mp t_{kh,j;me,j} )$ 
(the minus sign is used for charge currents).
This can be generalized to any form of $G_0$, and so not only non-ideal
leads but arbitrary combinations of random and non-random scattering
can be treated (the simplification due to the polar decomposition does not
easily generalize, however).

The results are summarized in Table~1, using Nazarov's notation.
$g_{k,m}=\sum_{j=1}^{W_{k,m}} T_j$ denotes the dimensionless
conductance of the lead, in the absence of proximity effects.
The Andreev amplitudes $f_k$ are related to Nazarov's 
``spectral vectors'' $\hat s_k$ by a stereographic projection from the
complex unit disc onto the upper hemisphere:
$f_k = \tan(\vartheta_k/2) \exp(i\varphi_k)$ where $\vartheta_k$ and
$\varphi_k$ are the polar and azimuthal angle of $\hat s_k$.
The angle difference $\Delta \theta_{k,m}$ is the ``distance'' between
$\hat s_k$ and $\hat s_m$ on the sphere, given by
$\hat s_k \cdot \hat s_m = \cos(\Delta \theta_{k,m})$, or 
$\tan (\Delta \theta_{k,m}/2) = |\beta_{k,m}|$.
In the last two lines of the table, we specialize the general
expressions to the cases treated in Ref.~\cite{Nazarov}
--- diffusive leads and leads containing strong tunnel barriers 
($T_j \to 0$) --- reproducing the older results. 
The complex phases of the $J_{k,m}$s are also simply related to the
directions of the spectral currents in Nazarov's picture, given by 
$\hat s_k \times \hat s_m$.  

The results for (the electrical conductance of) diffusive leads can be
obtained by two different methods: First, by integrating over the
known distribution of transmission eigenvalues in such leads
\cite{integ}, one finds that the enhancement of electrical
conductance by Andreev reflections in relatively open channels 
($T_j \s\to 1$) is exactly counterbalanced by the inhibition of 
``Cooper-pair tunneling'' in relatively closed channels ($T_j \s\to 0$), 
regardless of the value of $\Delta \theta_{k,m}$; Second, by treating 
the diffusive lead as $L \gg 1$ ideal (or tunnel) leads connected in
series through $L-1$ additional junctions, one finds that the spectral 
vectors of the intermediate junctions lie on the big arc that connects
$\hat s_k$ and $\hat s_m$ with angle differences of 
$\Delta \theta_{k,m} /L$, resulting in very small proximity induced
corrections to electrical conductivity ($L$ occurences
of a quantity which is quadratic in $\Delta \theta_{k,m} /L$).
Thus, electrical conductance in diffusive leads is unchanged by the
proximity effects being considered (see Fig.~2b below).

Motivated in part by the ``Andreev interferometer'' experiment of 
Petrashov {\it et al.} \cite{Petrashov}, we plot in Fig.~2 the electric
and heat conductances of three structures of a 
geometry shown in the inset: one with ideal leads, one with diffusive
leads, and one with tunnel barriers (each structure has 8 identical
leads).  The results are obtained numerically by repeatedly shifting
each of the $f_k$s in the complex plane by a fraction of the mismatch
in the spectral current, $\sum_m J_{k,m}$, until the iterations
converge.  An applied magnetic field leads to fast oscillations due to
the induced phase-difference between the two superconducting
electrodes, and slow oscillations (only one period shown) due to the
Aharonov Bohm flux threaded through a small ring in the structure.
The results exhibit some similarities to the experimental data, but as
stated above, we are unable to reproduce the experimental oscillations
(of the order of 10\%) of the electrical resistance in a diffusive
system.  As pointed out in Ref.~\cite{Stoof}, these may be due to the 
finite temperature involved (20 mK), requiring a generalization of the
present analysis.  

In summary, using diagrammatic techniques of random
scattering-matrix theory, we have rederived and enhanced Nazarov's
circuit theory of Andreev conductance.  The results follow in fact 
quite simply from notions of multiple Andreev scattering, which is
taken to occur locally in the $k$th junction with an amplitude $f_k$
(this is the amplitude of the ``giant Andreev backscattering peak''
\cite{Been_GBP}).  The values of $f_k$ are found self-consistently in
a manner analogous to finding the voltages in the junctions of an
electrical circuit.  Heat transport was also briefly considered here,
whereas the extension to finite temperatures and voltages is an
important goal for future developments.

I would like to thank C.W.J. Beenakker, P.W. Brouwer, B. Spivak and
A. Zee for numerous discussions.  Financial support of a Fulbright
fellowship and NSF grants No.\ PHY94-07194 and No.\ DMR93-08011 is
acknowledged.

\vskip 6cm
\newpage

\centerline{\large  TABLE}
\vskip 1cm

{\small\noindent
\begin{tabular}{c|ccc}

{\normalsize type of lead ~} & {\normalsize spectral current}   & 
  {\normalsize electrical conductance}  & {\normalsize heat conductance} \\
\cline{1-4}
ideal ($T_j \s= 1$) ~ & $W_{k,m} 2 \tan \half \Delta \theta_{k,m}$
             & $W_{k,m} \cos^{-2} \half \Delta \theta_{k,m}$ 
 & ~ $W_{k,m} { \mbox{$\cos \vartheta_k \cos \vartheta_{m \vphantom{_a}} 
\vphantom{A^{A^{A^A}}} $}
   \over \mbox{$ \cos^{2\vphantom{^a}} \half \Delta \theta_{k,m} $} } $ \\
general ~ &  $ \>\> \sum_j 
{ \mbox{$ T_j \sin \Delta \theta_{k,m} $} \over
  \mbox{$ 1 -T_j \sin^{2\vphantom{^a}} \half \Delta \theta_{k,m}$} } \;$ 
   & $ \;\; \sum_j T_j
{\mbox{$ \cos\Delta\theta_{k,m} +T_j \sin^2 \half \Delta \theta_{k,m}$} \over
 \mbox{$ (1 - T_j \sin^2 \half \Delta \theta_{k,m})^{2\vphantom{^a}}$} }$  
&  $ \;\; \sum_j 
{ \mbox{$ T_j \cos \vartheta_k \cos \vartheta_m $} \over 
 \mbox{$ 1 - T_j \sin^{2\vphantom{^a}} \half \Delta \theta_{k,m} $} }
\vphantom{\pmatrix{A^A \cr A_j}} $ \\
tunneling ($T_j \s\to 0$)~ &  $g_{k,m} \> \sin \Delta \theta_{k,m}$ 
                   & $g_{k,m} \cos \Delta \theta_{k,m}$
         &  $g_{k,m} \cos \vartheta_k \cos \vartheta_m 
\vphantom{{A \over A}} $ \\ 
diffusive ~ & $g_{k,m} \> \Delta \theta_{k,m}$
                     & $g_{k,m}$
          & $g_{k,m} 
{ \mbox{$\cos \vartheta_k \cos \vartheta_m \, \Delta \theta_{k,m} $}
      \over \mbox{$ \sin \Delta \theta_{k,m} \vphantom{^a}$} }
\vphantom{\pmatrix{\half \cr 0}} $ \\
\end{tabular}
}

\medskip

\indent\hspace*{21pt}\parbox{14cm}{ Table~1:
The dimensionless spectral current, $|J_{k,m}|$, and electric and thermal
conductances, $I_{k,m}/(V_m \s- V_k)$ and $Q_{k,m}/(U_m \s- U_k)$, 
for various types of leads.
}


\begin{figure}[h]
\vspace{0.2cm}
\epsfxsize=16cm
\epsffile{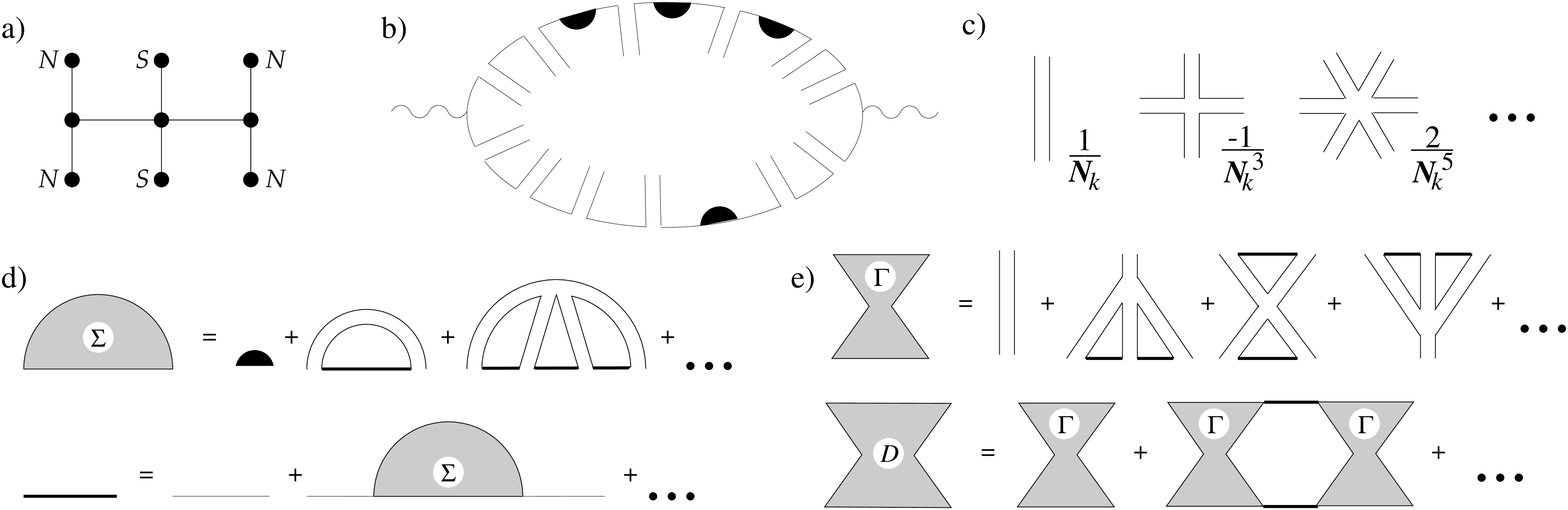}
\vspace{0.5cm}
\noindent\parbox{16cm}{Fig.~1:
Random-scattering-matrix diagrams for Andreev circuit theory 
(following Ref.~\cite{AZ}).  a) Schematic drawing of a circuit with 
$K \s= 3$ junctions, and six external electrodes ($M \s= 9$), two
of which are superconducting.  b) Typical conductance diagram, before
averaging; single light lines represent ``deterministic'' transport in
the leads ($G_0$), dangling double lines represent ``random''
scattering in the junctions, and full semicircles represent Andreev
scattering at the superconducting electrodes.  c) Couplings used to
join the dangling lines upon averaging, and their weights.
d) Diagrams for the ``one-particle irreducible part'' 
$\Sigma$, and averaged propagator $G$ (thick line).  e) Diagrams for the 
``two-particle irreducible part'', $\Gamma$, and the ``diffuson'' $D$.
}
\end{figure}

\begin{figure}[h]
\vspace{0.5cm}
\epsfxsize=16cm
\epsffile{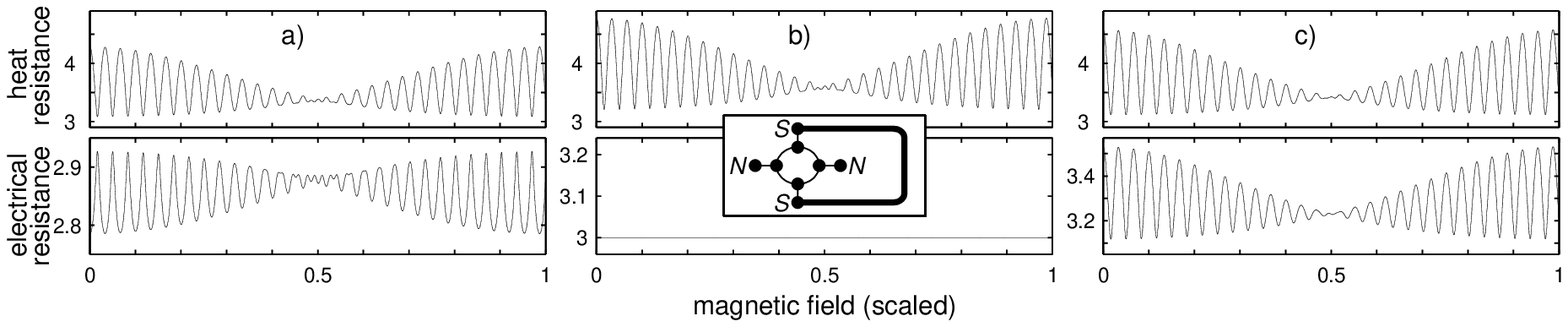}
\vspace{0.2cm}
\noindent\parbox{16cm}{ Fig.~2:
Overall electrical and heat resistances between the two normal (N)
electrodes of the circuit shown in the inset, as a function of magnetic
field, in units of the resistance of a single lead (resistance in
the absence of proximity effects $= 3$).  a) Ideal leads; b)
diffusive leads; c) leads with tunnel barriers.  The area enclosed by
the normal loop (thin circle) is some 30 times smaller than that of the
superconducting loop (thick line in inset).
}
\end{figure}


\begin{references}
\bibitem{rem} The conductances can be derived solely from the
block off-diagonal terms in \Eq{vGtv}.  The consistency with the
diagonal terms (i.e.\ current conservation in the leads) relies 
on the vanishing of $f_k^* \sum_m J_{k,m} + {\rm c.c.}$.
By reversing the argument one may derive the expression
for $\Gamma$, or prove that the component of $J_{k,m}$
parallel to $f_k$ is conserved.
\bibitem{imbalance} The voltages $V_m$ here are the electro-chemical
potentials of the quasiparticles; the electrostatic potentials are
obtained by appropriately weighting this with the (vanishing)
electro-chemical potential of the superconducting component (see
Ref.~\cite{Stoof} below).
\end{references}
\end{document}